\documentclass[fleqn,twoside]{article}
 \usepackage[headings]{espcrc2}

\usepackage[dvips]{graphicx}
\usepackage{amsmath,amssymb}

\title{\hfill \hbox{RUB-TPII-10/09}\vspace*{5mm}\\
       Nonlocal condensates and pion form factor
                \thanks{Presented by the first author at the 3rd Joint International Hadron Structure'09 Conference, 
                 Tatranska Strba (Slovak Republic), Aug. 30--Sept. 3, 2009
                 }
       }

\author{A.~V.~Pimikov\address[BLTP]{Bogoliubov Laboratory of Theoretical Physics,
                 JINR, 141980 Dubna, Moscow region, Russia},
        A.~P.~Bakulev\addressmark[BLTP],
        N.~G.~Stefanis\address{Institut f\"{u}r Theoretische Physik II,
                 Ruhr-Universit\"{a}t Bochum, D-44780 Bochum, Germany}
      }

\runtitle{Nonlocal condensates and pion form factor}
\runauthor{A.~V.~Pimikov et al.}

\begin{document}

\begin{abstract}
We review a nonlocal-condensate approach and its application
in QCD sum rules for the spacelike electromagnetic pion form factor.
It is shown that the nonlocality of the condensates is a key point
to include nonperturbative contributions to the pion form factor.
\vspace{1pc}
\end{abstract}

\maketitle

The spacelike electromagnetic pion form factor (FF) describes the
scattering of charged particles off the pion by exchanging a photon.
It is defined by the following matrix element:
\begin{eqnarray*}
  \langle{\pi^{+}(P^{\prime})| J_{\mu}(0) | \pi^{+}(P)\rangle}
    =  {\left( P + P^{\prime}\right)}_{\mu} F_{\pi}(Q^{2}) \,,
\end{eqnarray*}
where $J_\mu$ is the electromagnetic current
and $q$ is the photon momentum $q^2=(P^{\prime} - P)^2=-Q^2<0$
in the spacelike region.
For asymptotically large momenta, one can apply the factorization 
theorem so that the FF will be represented via the inverse moment of 
the leading-twist pion distribution amplitude (DA).
The precise value of $Q^2$ at which this perturbative term
starts to prevail cannot be determined accurately.
The estimates for the crossover momentum scale range from
$100$~GeV$^2$ \cite{IL84,JK93,BLM07}
down to values around
$20$~GeV$^2$ \cite{SSK,BPSS04}.
But even this latter relatively small momentum is hopelessly far away
from the capabilities of any operating or planned accelerator
facility.

\section{QCD sum rules approach}

At intermediate momentum transfers factorization fails;
therefore one needs to apply nonperturbative approaches.
One of these methods was suggested in 1979 by Shifman, Vainshtein, and 
Zakharov~\cite{SVZ} and was called QCD Sum Rules (SR)s.
To extract information about the pion form factor in the QCD SR 
approach, one needs to investigate the Axial-Axial-Vector (AAV) 
correlator of three currents:
\begin{eqnarray*}
    \int\!\!\!\!\!\int\!\!d^4x\,d^4y\,e^{i(qx-P^\prime y)}
    \langle{0|T\!\!\left[J^{+}_{5\beta}(y) J^{\mu}(x) J_{5\alpha}(0)\right]\!\!|0\rangle} \ ,
\end{eqnarray*}
where
$J^{\mu}(x)=   e_u\,\overline{u}(x)\gamma^\mu u(x)
             + e_d\,\overline{d}(x)\gamma^\mu d(x)$
is the electromagnetic current current and
$J_{5\alpha}(x)=   \overline{d}(x)\gamma_5\gamma_\alpha u(x)$
is the axial-vector current.
For simplicity, let us describe how this method works using as an 
example the two-point correlator
\begin{eqnarray*}
 \Pi(Q^2)=
     \int\!d^4x\,e^{iqx}
     \langle{0|T\!\left[J(x) J(0)\right]\!|0\rangle}\,.
\end{eqnarray*}
There are two ways to calculate this correlator.
The first one is based on the dispersion relation
\begin{eqnarray*}
 \Pi_\text{had}\left(Q^2\right)
  = \int\limits_{0}^\infty \frac{\rho_\text{had}\!\left(s\right) ds}{s+Q^2}
  + \text{subtractions}\, ,
\end{eqnarray*}
where physical observables (masses $m_h$ and decay constants $f_h$)
can be introduced in the calculation by a model spectral density
as the sum of the first resonance contribution plus the contribution 
of the continuum beginning at the threshold $s_0$:
$$\rho_\text{had}\!\left(s\right)
 = f_h^2\,\delta\left(s-m_h^2\right)
 + \rho_\text{pert}\!\left(s\right)\theta\left(s-s_0\right)\,.$$
Higher states are taken into account by the perturbative spectral 
density $\rho_\text{pert}\!\left(s\right)$
on account of the quark-hadron duality.
The second approach employs the operator product expansion (OPE):
$$\Pi_\text{OPE}\left(Q^2\right)
 = \Pi_\text{pert}\left(Q^2\right)
 + \sum\limits_n C_n \frac{\langle{0|:O_n:|\,0\rangle}}{Q^{2n}}\,.
$$
Vacuum expectation values $\langle{O_n\rangle}$
\footnote{\small Hereafter we write $\langle{O_n\rangle}\equiv\langle{0|:O_n:\!|\,0\rangle}$.}
of the normal product of quark and gluon fields
are not vanishing but constitute (nonperturbative) condensates.
Demanding the agreement between the results of these two calculations,
we obtain the following SR
\begin{eqnarray*}
\Pi_\text{had}\left(Q^2,m_{h},f_{h}\right)
       =
\Pi_\text{OPE}\left(Q^2            \right)
\end{eqnarray*}
that allows us to extract the introduced hadronic parameters---masses 
and decay constants in the case of the two-point SR---from the 
condensates $\langle{O_n\rangle}$.
On the other hand, the three-point SR~\cite{NR82,IS82} helps us
to study hadronic form factors.

The simplest condensate is the so-called quark condensate.
Consider the vacuum expectation value of the $T$-product of two quark 
fields:
\begin{eqnarray} \label{eq:Tqq}
 \langle{0|T\left(\bar{q}_B(0)q_A(x)\right)|0\rangle}
 ~~~~~~~~~~~~~~~~~~~~~
\\ \nonumber
 = \langle{0|:\bar q_B(0) q_A(x):|0\rangle} -i \hat{S}_{AB}(x)
 \,,~~~~~
\end{eqnarray}
where $A,~B$ are Dirac indices.
The second term here corresponds to the usual propagator due to the 
Wick theorem, while the first one is the quark condensate.
From this equation one can see that the quark condensate is an 
additional nonperturbative contribution to the quark propagator.

In perturbation theory the vacuum coincides with the ground state
of the free-field theory; hence the expectation value of the normal 
product is zero.
Therefore, there are no condensate terms in perturbation theory.
However in the physical vacuum this is not the case.
For this reason, in the standard QCD SR approach, the nonzero quark 
condensate 
$\langle{\bar q q\rangle}\equiv\langle{\bar q_A(0) q_A(0)\rangle}$
appears.
The value of this constant was defined through comparison with 
experimental data for the $J/\psi$-meson~\cite{SVZ}.
Assuming a small coordinate dependence, the quark condensate can be 
represented by the first two terms of the Taylor expansion:
\begin{eqnarray}
\label{eq:loc.cond}
 \langle{\bar q_B(0)\,q_A(x)\rangle}
  = \frac{\delta_{AB}}{4}\,
     \bigg[ \langle{\bar q q\rangle}
          + \ldots
     \bigg]~~~~~~~~~~~~~
  \\\nonumber
 ~~~~~~~+
   i\,\frac{\widehat{x}_{AB}}{4}\,
       \frac{x^2}{4}\,
        \bigg[\frac{2\alpha_s\pi\langle{\bar qq}\rangle^2}{81}+ \ldots
        \bigg]
   \,,
\end{eqnarray}
where we kept the scalar and vector parts apart.
Note that the condensates in this representation are local.

\section{Pion FF in the QCD SR approach}

Unfortunately, the local approximation (\ref{eq:loc.cond}) is not 
reasonable for studying form factors (FF) and distribution amplitudes, 
as it was stated in~\cite{MR-NLC,BR91,MS93,BPS09}.
The reason is the unphysical behavior of the local condensate 
(\ref{eq:loc.cond}) at large $x^2$, entailing a constant scalar and a 
vector part that is even growing with the distance between the quarks 
$x^2$.
As a result, the nonperturbative part of the OPE linearly increases
with the momentum $Q^2$ in the case of the FF (or with the moment $N$ 
in the case of the DA):
$\left(c_1 + Q^2/M^2\right)$,
where $c_1$ is a dimensionless constant (not depending on $Q^2$)
and $M^2$ is the Borel parameter.
At the same time, the perturbative part decreases, while the
nonperturbative one increases with $Q^2$, hence generating an 
inconsistency of the SR at intermediate and large $Q^2$.
Therefore, we can not rely upon the obtained SR for the pion FF
for momentum values $Q^2>3$~GeV$^2$~\cite{BPS09}.

In order to improve the $Q^2$ dependence, one needs to modify the model
of the quark-condensate behavior at large distances.
Indeed, lattice simulations~\cite{DDM99,BM02} and instanton models
\cite{DEM97,PW96} indicate a decreasing of the scalar quark condensate
with increasing interquark distance, thus confirming the approach of 
nonlocal condensates (NLC)s \cite{MR-NLC}.
To further improve the condensate contribution, one may calculate terms
which contain higher-dimension operators of the form
   $\langle \bar q(0)D^2q(0)\rangle$,
   $\langle \bar q(0)(D^2)^2q(0)\rangle$,
etc., originating from the Taylor expansion of the original nonlocal 
condensate, i.e.,
$\langle{\bar q_B(0)\,q_A(x)\rangle}$.
The resulting total condensate contribution decreases for large $Q^2$.
However, each term of the standard OPE has the structure $(Q^2/M^2)^n$,
and one should, therefore, resum them to get a meaningful result.
The main strategy of the NLC SR~\cite{MR-NLC,BR91,MS93} is to avoid 
the original Taylor expansion and deal directly with the NLCs by 
introducing model functions that describe the coordinate dependence 
of the condensates.

In the NLC approach the bilocal quark-antiquark condensate
has the following form
(we use the Euclidean interval $x^2 = -x_0^2-\vec{x}^2<0$):
\begin{eqnarray}
 \langle{\bar{q}_A(0) q_B(x)}\rangle
  = \frac{\delta_{BA} \langle{\bar{ q} q}\rangle}{4}
     \int_0^\infty\!\! f_S(\alpha)
      e^{\alpha x^2/4}\,d\alpha
  \\ \nonumber
  - \frac{iA_0\,\widehat{x}_{BA}}{4}
     \int_0^\infty \!\!f_V(\alpha)
      e^{\alpha x^2/4}\,d\alpha\,,
\end{eqnarray}
which, for the most general case, is parameterized by the distribution 
functions
$f_S(\alpha)$ and $f_V(\alpha)$,
with $A_0=2\alpha_s\pi\langle{\bar{q}q}\rangle^2/81$.
As usual in the QCD SR approach, the fixed-point (Fock--Schwinger) 
gauge $x^\mu A_\mu(x)=0$ is used.
For this reason, all connectors
${\mathcal C}(x,0) \equiv
 {\mathcal P}
  \exp\!\left[-ig_s\!\!\int_0^x t^{a} A_\mu^{a}(y)dy^\mu\right]=1
$,
are evaluated along a straight-line contour going from $0$ to $x$.
The explicit form of these functions must be taken from a concrete 
model of the nonperturbative QCD vacuum derived either from  the 
exact solution of QCD or by applying some approximation, e.g.,
a QCD simulation on the lattice or the employment of the instanton 
approach.
In the absence of information on the coordinate dependence of the 
quark condensate, it was proposed~\cite{MR-NLC} to use the first 
nontrivial approximation which takes into account only the finite 
width of the spatial distribution of the vacuum quarks:
$f_S(\alpha)
  = \delta\left(\alpha-\lambda_q^2/2\right)
$.
This generates a Gaussian form of the NLC in the coordinate 
representation:
$\langle{\bar{q}_A(0) q_A(x)}\rangle 
= 
\langle{\bar{q}q}\rangle e^{-|x|^2\lambda_q^2/8}
$.
The Gaussian form leads to the following form of the condensate 
contributions to the FF:
$\left(c_1 + Q^2/M^2\right)\,e^{- c_2 Q^2\lambda_q^2/M^4}$,
where $c_i$ are dimensionless constants (not depending on $Q^2$).
One can see that taking into account the nonlocality of the condensates
($\lambda_q^2\neq0$), enables us to obtain a decreasing behavior of the 
nonperturbative part of the FF at large $Q^2$.

The same techniques should be applied to deal also with the mixed 
quark-gluon condensate:
$\langle{\bar{q}_B(0)(-g A^a_\nu(y)\,t^a)q_A(x)}\rangle$.
There are two models for this condensate: the minimal and the improved 
one.
The explicit form of these models can be found in~\cite{BP06,BPS09}.
The nonlocal gluon-condensate contribution produces a very complicated 
expression.
But owing to its smallness, we can model the nonlocality of the 
gluon-condensate in analogy to the quark case using an exponential 
factor~\cite{BR91,MS93}, viz.,
$e^{-\lambda_g^2 Q^2/M^4}$.

\begin{figure}[h]
 \centerline{\includegraphics[width=0.45\textwidth]{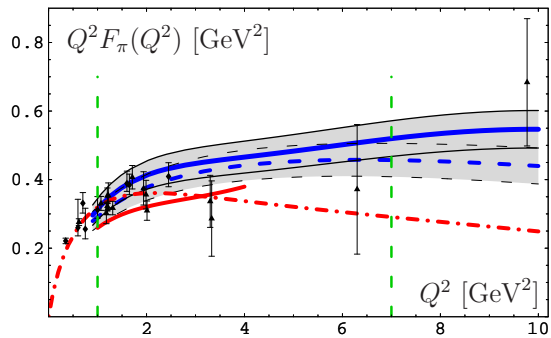}}
  \caption{\label{fig:Q2FF}
   Scaled pion form factor $Q^2 F_{\pi}(Q^2)$ for the minimal NLC model
   (shown as a thick broken line inside the shaded band delimited by the dashed lines
   which denote the uncertainty range).
   The improved NLC model is represented by a solid line inside the shaded band within
   the solid lines ($\lambda_q^2=0.4$~GeV$^2$).
   The short solid line at low $Q^2$ shows the result of the
   standard QCD SR with local condensates \cite{NR82,IS82}.
   The dashed-dotted line denotes the estimate derived in \cite{BLM07}
   with Local Duality QCD SRs.
   The two broken vertical lines mark the region, where the influence
   of the particular Gaussian model used to parameterize the QCD vacuum
   structure in the NLC QCD SRs is not so strong.}
\end{figure}

The described NLC QCD SR approach provides the basis of the theoretical 
framework for the calculation of the pion form factor proposed in our 
recent paper \cite{BPS09}.
This method yields predictions for the spacelike pion form factor 
(see Fig.~\ref{fig:Q2FF}) that compare well with the experimental data
of the Cornell \cite{FFPI-Cornell} (triangles)
and the JLab Collaborations \cite{JLab08II} (diamonds)
in the momentum region currently accessible to experiment.
These predictions cover also the range of momenta to be probed by the 
$12$~GeV$^2$ upgraded CEBAF accelerator at the Jefferson Lab in the 
near future.
This planned high-precision measurement of the pion FF at JLab
will certainly help to check the quality of the discussed NLC models.

\section{Conclusions}
We presented an analysis of the spacelike pion form factor using the 
SR in connection with two different models for the nonlocal condensate.
The local condensates used in the pioneering studies~\cite{NR82,IS82}
lead to an unphysical increasing of the nonperturbative contribution
to the pion form factor at large $Q^2$, the reason being the unnatural 
behavior of the local condensates (\ref{eq:loc.cond}) at large $x^2$.
This behavior restricts the region of the QCD SR applicability to 
$Q^2\lesssim3~\text{GeV}^2$.
In contrast to the local condensates, the nonlocal ones decrease at 
large $x^2$, hence inducing the decay of the nonperturbative terms at 
large $Q^2$.
This makes the QCD SRs stable and enlarges the region of its 
applicability towards momenta as high as $10~\text{GeV}^2$.

\section{Acknowledgments}
This work was supported in part by
the Russian Foundation for Fundamental Research,
grants No.\ ü~07-02-91557, 08-01-00686, and 09-02-01149,
the BRFBR--JINR Cooperation Programme,
contract No.\ F08D-001,
the Deutsche Forschungsgemeinschaft
(Project DFG 436 RUS 113/881/0-1),
and
the Heisenberg--Landau Program under grant 2009.
A.~V.~P. acknowledges support from the Program
``Development of Scientific Potential in Higher Schools''
(projects 2.2.1.1/1483, 2.1.1/1539).


\end{document}